\documentstyle[preprint,aps,prb,epsfig]{revtex}
\begin{document}

\title{Highly oriented VO$_{2}$ thin films prepared by sol-gel deposition method}

\author{Byung-Gyu Chae, Hyun-Tak Kim, Sun-Jin Yun, Bong-Jun Kim, Yong-Wook Lee, Doo-Hyeb Youn, and Kwang-Yong Kang}

\address{Basic Research Laboratory, ETRI, Daejeon 305-350,
Republic of Korea}

\maketitle{}

\begin{abstract}
  Highly oriented VO$_{2}$ thin films were grown on sapphire
substrates by the sol-gel method that includes a low pressure
annealing in an oxygen atmosphere. This reduction process
effectively promotes the formation of the VO$_{2}$ phase over a
relatively wide range of pressures below 100 mTorr and
temperatures above 400$^{\circ}$C. X-ray diffraction analysis
showed that as-deposited films crystallize directly to the
VO$_{2}$ phase without passing through intermediate phases.
VO$_{2}$ films have been found to be with [100]- and
[010]-preferred orientations on Al$_{2}$O$_{3}$(\=1012) and
Al$_{2}$O$_{3}$(10\=10) substrates, respectively. Both films
undergo a metal-insulator transition with an abrupt change in
resistance, with different transition behaviors observed for the
differently oriented films. For the [010]-oriented VO$_{2}$ films
a larger change in resistance of 1.2$\times$10$^{4}$ and a lower
transition temperature are found compared to the values obtained
for the [100]-oriented films.

\end{abstract}
\pacs{ }

  Vanadium dioxide (VO$_{2}$) has been known to undergo the metal-insulator
transition (MIT) near 68$^{\circ}$C accompanied by abrupt changes
in electrical resistivity and infrared transmission.\cite{1,2,3,4}
These transition properties make VO$_{2}$ films suitable for
technological applications such as electrical switches and
electro-optical modulators.\cite{3,5} VO$_{2}$ films have shown
reproducible switching characteristics without sample degradation,
while bulk materials did not endure repeated transitions due to
stress. Although films with transition characteristics comparable
to single crystal have been grown, there have not yet been any
representative devices realized using this system. One of the
reasons may be the stabilization of VO$_{2}$ phase. The formation
of VO$_{2}$ phase is complicated due to the fact that there are
several phases in vanadium oxide system with oxygen stoichiometry.
Establishing a process for the fabrication of VO$_{2}$ film that
possesses stable transition characteristics is crucial for the
realization of devices.

  Up to now, VO$_{2}$ films have been fabricated by various methods
such as pulsed laser deposition,\cite{6,7,8,9}
sputtering,\cite{10,11,12} and chemical vapor
deposition\cite{13,14}. Among these methods, researchers have
focused on the sol-gel method for depositing VO$_{2}$ films
because it has many advantages such as low cost, large area
deposition, and the feasibility of metal-doping.\cite{15,16,17}
Fabrication of films with a variety of transition properties is
needed in order to satisfy particular device specifications. The
switching behavior of the MIT strongly depends on the
crystallinity and the stoichiometry of the films. The sol-gel
method can easily make films with various stoichiometries and
metal dopants, thus permitting the growth of numerous films with a
specific transition behavior. Therefore, this method is suitable
for fabricating films for switching devices. However, the
formation of the pure phase is not easy even when employing the
sol-gel technique due to the difficulty in controlling the
annealing condition. The annealing regime of the films has been
known to be crucial for forming the VO$_{2}$ phase. It has been
reported that a reducing gas atmosphere at a low pressure is
required for successful phase formation. Moreover, vanadium oxide
system undergoes successive reduction with increasing annealing
temperature.\cite{15,17} This means that the films pass through
numerous intermediate phases during the annealing process, thus
resulting in a narrower fabrication window.

  In this research, we have successfully fabricated VO$_{2}$ films
by the sol-gel method with a more simplified annealing process.
The formation of the VO$_{2}$ phase was affected by annealing in
oxygen only without a reducing gas. The films have highly
preferred orientations related to the sapphire substrates. We
investigate the different characteristics of the transition
behavior with respect to the film orientation.

  VO$_{2}$ films were grown on sapphire substrates by a sol-gel
method. This method consists of spin-coating and a subsequent
annealing process under low pressure. The coating solutions were
prepared by synthesizing the vanadium tri-isopropoxide,
VO(OC$_{3}$H$_{7}$)$_{3}$ (Stream Ltd. USA), in isopropanol using
a catalyst of acetic acid. The final 0.12 M solution concentration
was thus prepared. The solution was spin-coated onto sapphire
substrates with a spin rate of 2000 rpm for 20 s. The coated films
were dried at 250$^{\circ}$C for 3 min on a hot plate to remove
the excess alcohol. The procedure was performed in air to
partially hydrolyze the alkoxide film with ambient moisture and
repeated three times in the same manner. The thickness per cycle
was approximately 32 nm. The final heat treatment was carried out
at 410$^{\circ}$C for 30 min in air, which made the films turn
orange-yellow in color. In order to form the VO$_{2}$ phase
through a reduction process, subsequent annealing of the films was
done in a low pressure of oxygen.

  An X-ray diffractometer (XRD) using Cu K$\alpha$
was used to determine the film orientation. The surface morphology
and the grain structure of the films were investigated by scanning
electron microscopy (SEM). The electrical resistance of the films
was measured by using the four-probe method.

  Figure 1 shows X-ray diffraction patterns of the films annealed
at various temperatures. The films were annealed at each
temperature for 10 min at a pressure of 32 mTorr. The XRD
$\theta$-2$\theta$ scans indicate that both films have a strong
orientation relation to the sapphire substrate. The films on
Al$_{2}$O$_{3}$(\=1012) are grown with a [100]-preferred
direction, as shown in Fig. 1(a). As-deposited film without an
annealing has V$_{2}$O$_{5}$(001) peak, but its intensity is very
weak compared to the completely crystallized films on
sapphire.\cite{6} This indicates that the degree of the film
crystallization to the V$_{2}$O$_{5}$ phase is low. For the films
annealed over and at 400$^{\circ}$C, the strong XRD peak related
to the crystallization appears at a 2$\theta$ value of
37$^{\circ}$, which corresponds to the VO$_{2}$(200) reflection of
the monoclinic phase. Although the small peak with other direction
exists, we can confirm that the films have a strong orientation
from the comparative analysis of the peaks. The crystallization
behavior of the films grown on Al$_2$O$_3$(10\=10) is also similar
to that on (\=1012) except that the [010]-preferred orientation is
obtained, as shown in Fig. 1(b). The films have a strong (040)
peak of the monoclinic VO$_{2}$ phase. We also found that the
phase formation of the films during the annealing process depends
largely on the annealing temperature rather than the pressure. A
stable VO$_{2}$ phase is formed in the range from a few mTorr to
near 100 mTorr.

  The fact that no other peak expected for VO$_{2}$ phase is observed
in all films annealed at various temperatures is of particular
interest. As-deposited films have been known to undergo a
reduction process to the final VO$_{2}$ phase passing through
intermediate phases such as V$_{3}$O$_{7}$ and V$_{6}$O$_{13}$
during the annealing process.\cite{15,17} That is, the reduction
of vanadium oxide successively occurs as the annealing temperature
increases. However, our results show that the annealing process
does not follow the successive reduction of the vanadium oxide and
thus no intermediate phases appear at the various annealing
temperatures. Generally, the annealing process was carried out
under reducing gas mixture as like H$_2$ and CO, whereas our
process was performed in oxygen only without reducing gases.
Although the further study is required to understand completely
the phase formation, our simplified reducing process can be the
origin of this phenomenon. Our fabrication process of VO$_{2}$
films has considerable advantages over those in previous
reports.\cite{15,17} Since various intermediate phases are not
generated during the fabrication process, there is a large process
window for the production of high quality film.

  Figure 2 shows SEM images of the films.
Optically, the color of the VO$_{2}$ films is golden-brown. The
change of surface morphology of both films grown on (\=1012) and
(10\=10) with annealing temperature is similar. As shown in Fig.
2, as-deposited films without a subsequent annealing process show
well-formed grains although the crystallinity is very weak in the
X-ray patterns of Fig. 1. Small grains are generated in the
interior of initial grains, as illustrated in films annealed at
400$^{\circ}$C. This image may be related to the formation of the
VO$_{2}$ phase. Well crystallized films of the VO$_{2}$ phase at
470$^{\circ}$C have grains which are roundish in shape and the
average grain size is estimated to be about 100 nm.

  Figure 3 shows the temperature dependence of the electrical resistance for both films.
The characteristics of the resistance of the films can be another
indicator of the film quality. The measurement was performed
through sweep-up and -down processes. Both films undergo a
metal-insulator transition with temperature. The film resistance
decreases exponentially with increasing temperature and then an
abrupt drop occurs at the critical temperature. It has been
reported that the increasing hole density leads to the breakdown
of the sub-gap and thus the transition into the metallic
state.\cite{3,18} As shown in Fig. 3, the change in resistance
gets to as large as 8.7$\times$10$^{3}$ and 1.2$\times$10$^{4}$
for the VO$_{2}$/Al$_2$O$_3$(\=1012) and the
VO$_{2}$/Al$_2$O$_3$(10\=10) films, respectively. These values are
comparable to those of the single crystal, which means that the
films are highly oriented and possesses good
stoichiometry.\cite{10} Hysteresis in the resistance also occurs
during the sweep-up and -down processes of temperature.

  The inset of Fig. 3 displays the change in resistance curves
for the different annealing temperatures. The films reveal
different MIT behaviors with respect to the crystallinity. The
onset and sharpness of the transition are characterized by the
plot of d$R$/d$T$ for [100]-oriented films as shown in Fig. 4(a).
The transition width defined as the full width at half maximum of
this peak is as low as 0.6$^{\circ}$C. The sharpness of the
transition has been known to be related to the degree of
misorientation between adjacent grains.\cite{10} Well-matched
grains permit an effective propagation of the metallic regions in
the films without additional energy loss. The VO$_{2}$ films grown
on (\=1012) have a lower transition width than 3.1$^{\circ}$C for
the films on (10\=10), which indicates that the former films have
higher in-plane alignment of grains.

  Figure 4(b) shows the change in the transition temperature
($T_{\rm{c}}$) and the magnitude of the resistance change ($\Delta
R$) for both of the films annealed at various temperatures.
$T_{\rm{c}}$ is defined as the peak value of the d$R$/d$T$ curve.
The resistance of the films annealed at 400$^{\circ}$C decreases
slowly with little change at a critical temperature. This
indicates that the films are not totally crystallized although the
crystallinity of VO$_{2}$ phase is confirmed from the XRD
patterns. Both of the films annealed above 430$^{\circ}$C display
good transition behaviors as evidenced by change in resistance but
exhibit different transition characteristics. As shown in Fig.
4(b), for the VO$_{2}$/Al$_2$O$_3$(\=1012) films, the transition
temperature of 65$^{\circ}$C almost doesn't change with annealing
temperature above 430$^{\circ}$C, while the resistance change
slightly increases from 4.8$\times$10$^{3}$ to
8.7$\times$10$^{3}$. For the VO$_{2}$/Al$_2$O$_3$(10\=10) films,
the transition temperature slightly changes from 63$^{\circ}$C to
60$^{\circ}$C, while the magnitude of the resistance change
remains of order of 1.2$\times$10$^{4}$ with increasing annealing
temperature.

  [010]-oriented VO$_{2}$ films show relatively higher $\Delta R$ and lower
$T_{\rm{c}}$ than [100]-oriented films. This is consistent with
previously reported results.\cite{9} The direction in which the
resistance was measured with respect to the film orientation could
influence the transition characteristics. A further study is
necessary to understand the different transition properties of
both films in detail.

  In conclusion, VO$_{2}$ phase was well-formed on sapphire
substrates by using a simplified annealing process. The VO$_{2}$
films were grown in a wide range of pressures from several mTorr
up to 100mTorr of oxygen atmosphere and at temperatures above
400$^{\circ}$C, without passing through any intermediate phases.
The VO$_{2}$ films had a highly preferred orientation, and showed
resistance changes as large as 1.2$\times$10$^{4}$ which is the
comparable property of the single crystal.\cite{10} Different
transition properties were obtained depending on the film
orientation. Our method with a simplified annealing process leads
to the effective growth of high quality VO$_{2}$ films.

\begin{figure}
\vspace{0.0cm}
\centerline{\epsfysize=22.0cm\epsfxsize=15.0cm\epsfbox{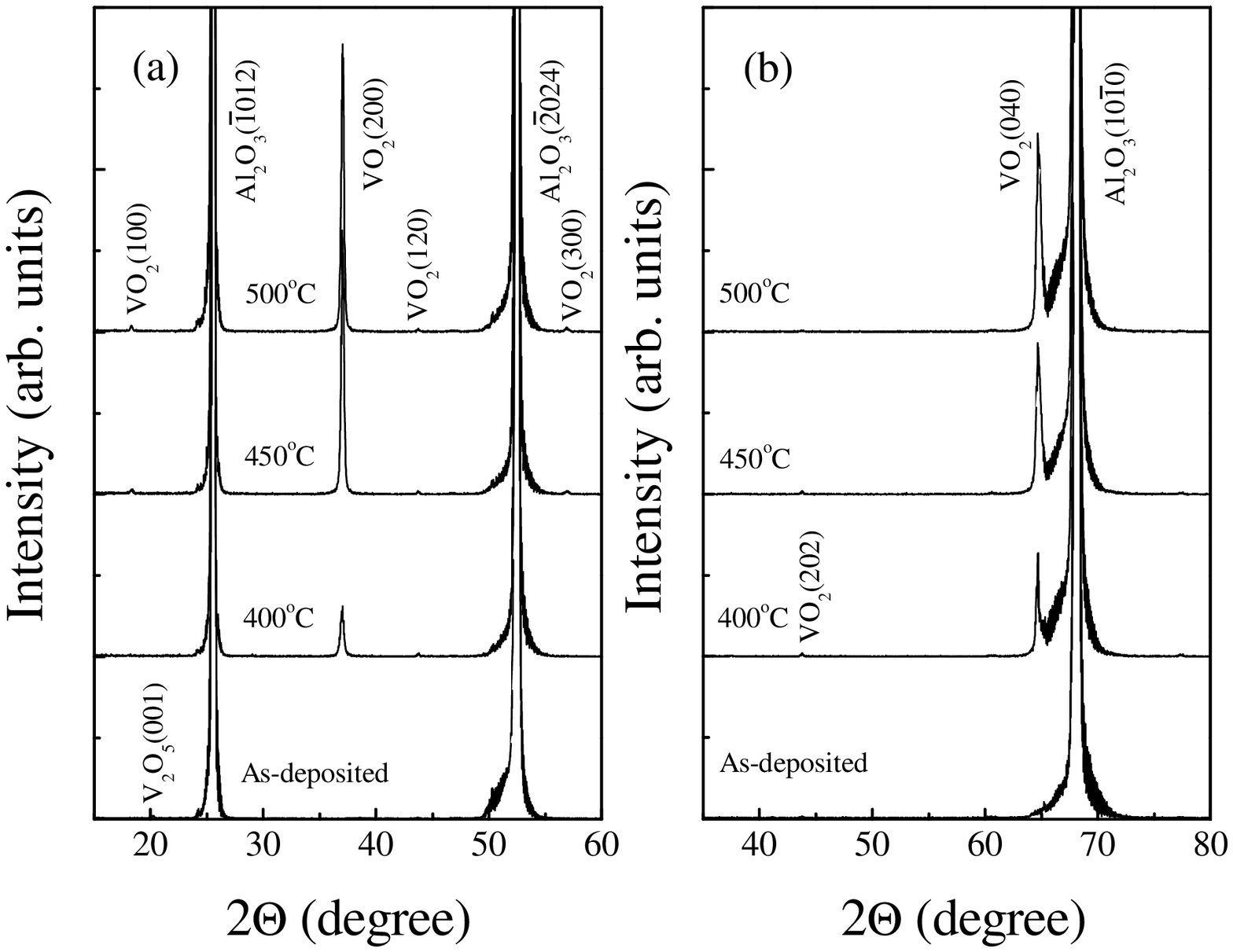}}
\vspace{-6.5cm} \caption{X-ray diffraction patterns of vanadium
oxide films grown on (a) Al$_{2}$O$_{3}$(\=1012) and (b)
Al$_{2}$O$_{3}$(10\=10) substrates. The films were annealed at
various temperatures in 32 mTorr of oxygen atmosphere.} \label{f1}
\end{figure}

\begin{figure}
\vspace{-5.0cm}
\centerline{\epsfysize=22.0cm\epsfxsize=15.0cm\epsfbox{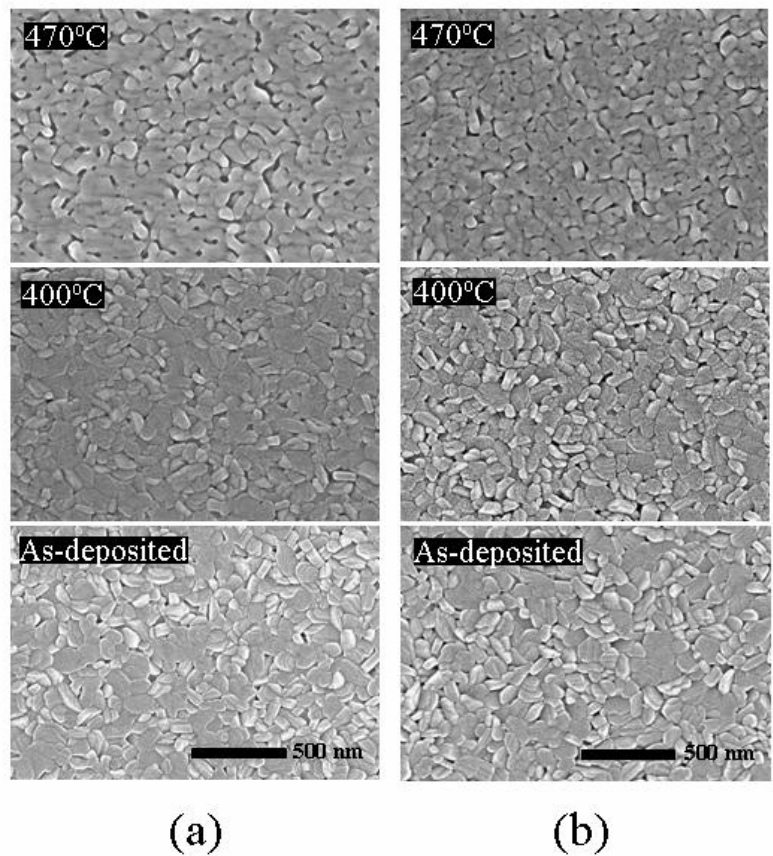}}
\vspace{-4.0cm} \caption{SEM images of vanadium oxide films grown
on (a) Al$_{2}$O$_{3}$(\=1012) and (b) Al$_{2}$O$_{3}$(10\=10)
substrates.} \label{f2}
\end{figure}

\begin{figure}
\vspace{-7.0cm}
\centerline{\epsfysize=22.0cm\epsfxsize=15.0cm\epsfbox{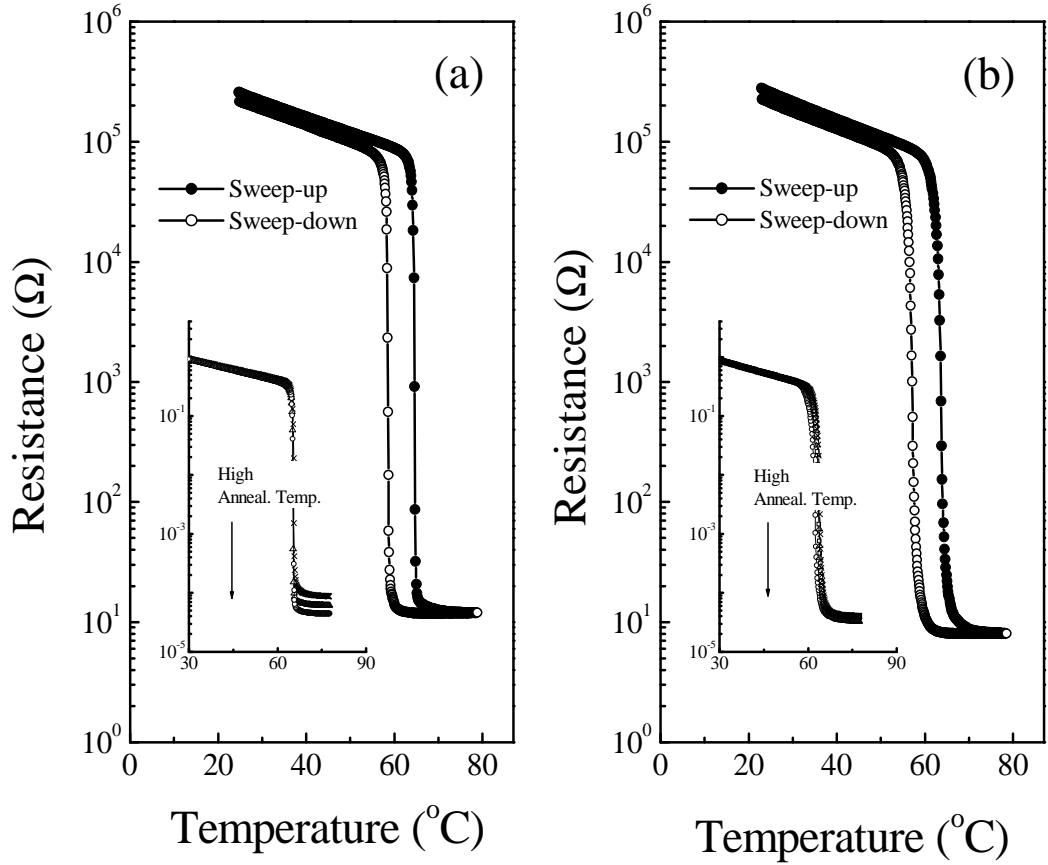}}
\vspace{-6.0cm} \caption{Changes in resistance for (a) the
VO$_{2}$/Al$_{2}$O$_{3}$(\=1012) and (b) the
VO$_{2}$/Al$_{2}$O$_{3}$(10\=10) films. The inset displays the
changes in resistance with annealing temperature with the curves
normalized for convenience.} \label{f3}
\end{figure}

\begin{figure}
\vspace{-5.0cm}
\centerline{\epsfysize=22.0cm\epsfxsize=15.0cm\epsfbox{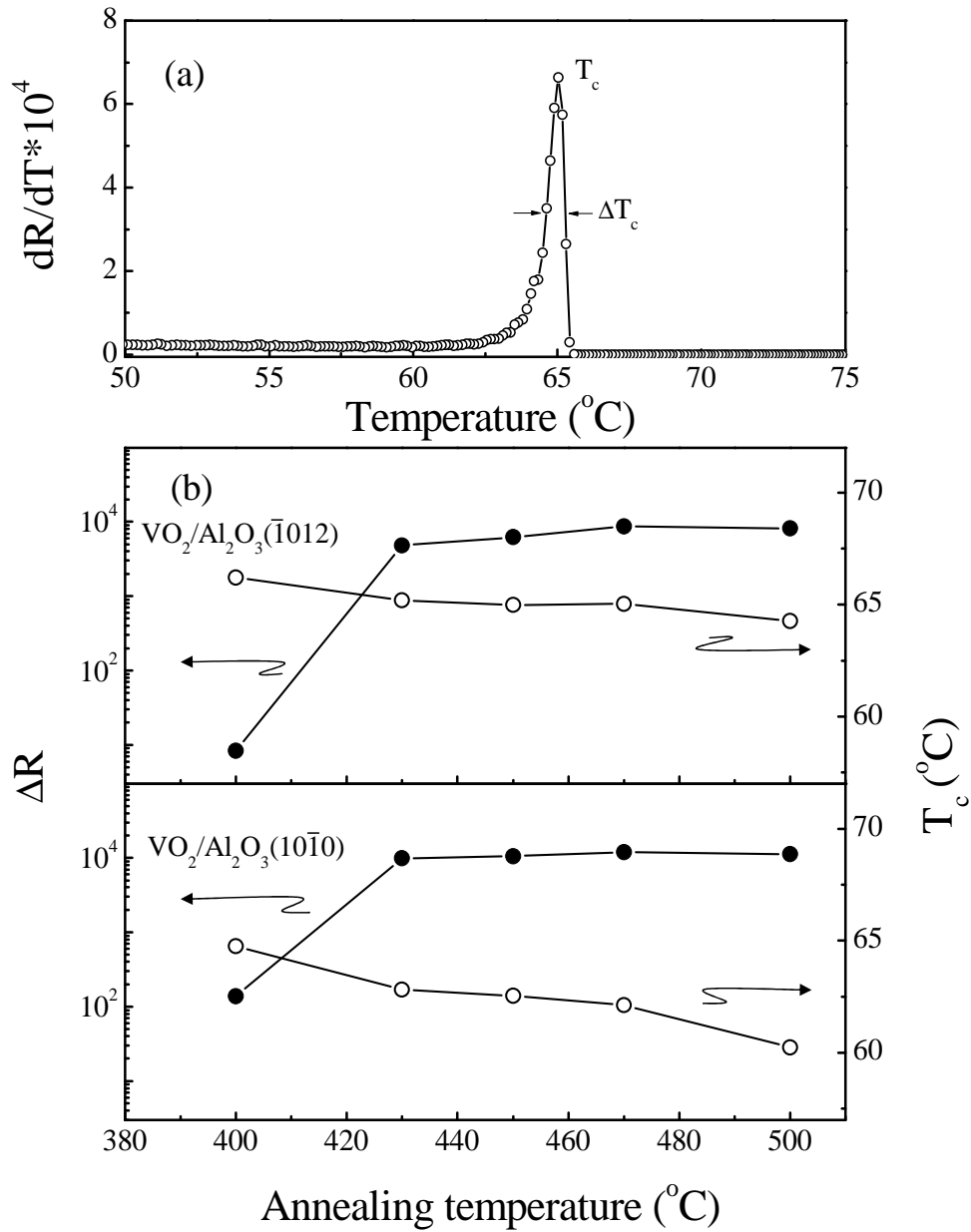}}
\vspace{-3.0cm} \caption{(a) Derivative of the change in the
resistance with respect to temperature for the film on
Al$_{2}$O$_{3}$(\=1012). (b) Changes in the magnitude of the
resistance change and transition temperature as a function of
annealing temperature for the films on Al$_{2}$O$_{3}$(\=1012) and
Al$_{2}$O$_{3}$(10\=10).} \label{f4}
\end{figure}

\end{document}